\documentclass[prd,showpacs]{revtex4}
\usepackage[dvipdfm]{hyperref,graphicx}
\usepackage{bm}
\begin{document}
\title{Problems of the Sensitivity Parameter and Its Relation to 
the Time-varying Fundamental Couplings Problems }

\author{Su Yan}
\email{yans@feynman.phys.northwestern.edu}
\affiliation{Department of Physics and Astronomy,
Northwestern University, Evanston, IL 60208, USA}
\begin{abstract}
The sensitivity parameter is widely used for quantifying 
fine tuning. However, examples show it fails to give correct results under certain circumstances.
We argue that these problems only occur when calculating the sensitivity of a dimensionful mass 
parameter at one energy scale 
to the variation of  a dimensionless coupling constant at another energy scale. 
Thus, by mechanisms such as dynamical symmetry breaking etc, the high sensitivity of the energy scale 
parameter \(\Lambda\) to the dimensionless coupling 
constant can affect the reliability of the sensitivity parameter through the renormalization invariant factor 
of the dimensionful parameter. Theoretically,
These phenomena are similar to the problems associated with the  
time-varying coupling constant discovered recently. We argue that, the reliability of the sensitivity parameter can 
be improved if it is used properly.

\end{abstract}
\pacs{11.10.Hi 12.10.Kt 12.60.Jv 14.80.Bn}

\maketitle

\section{Introduction}
In quantum field theories, as a consequence of quadratic divergent quantum corrections 
to the fundamental scalar masses, in order to obtain light weak scale observables,
delicate fine-tuning mechanisms are usually required. Due to the absence of reasonable
explanations to these fine-tuning mechanisms, quadratic divergent quantum corrections 
are usually problematic, which indicate the theories are incorrect. In light of this, 
Wilson and 't Hooft\cite{Wilson} introduced the principle of naturalness, which requires 
the radiative corrections to a measurable parameter should not be much greater
than the measurable parameter itself, thus a magical fine-tune mechanism 
is not required to have the theory agree with the current observations.

The simplest example of  such problems is the fundamental scalar of \(\phi^4\) model:

\begin{equation}
\mathcal{L}=\frac{1}{2}[(\partial_{\mu}\phi)^2-m^2_0\phi^2]-\frac{g}{4!}\phi^4
\label{eqx1}
\end{equation}

At one-loop the renormalization of the scalar mass is of the form:

\begin{equation}
m^2=m^2_0-g^2\Lambda^2
\label{eqx2}
\end{equation}
where \(m_0\) is the bare mass, \(\Lambda\) is the cut-off energy scale. 
Because of the tremendous scale difference between the light observable scalar mass and the 
bare mass, 
a fine-tuning mechanism that can adjust \(m_0\) and \(\Lambda\) 
very precisely is required to stabilize the observable mass. 
Otherwise any minute variations of \(m_0\) or \(\Lambda\) 
will completely change the value of the observable mass.

Motivated by the importance of such problems, find a way
to quantitatively describe the severity of fine-tuning is important. 
Think of a weak scale measurable parameter \(y\)
which is affected by the fine-tuning problem, it exhibits a strong dependence on 
a fundamental Lagrangian parameter \(x\) at the Planck scale. 
To calculate the severity of  fine-tuning for such instance, 
R. Barbieri and G.F.Giudice et al.\cite{BG} proposed the following sensitivity parameter:

\begin{equation}
c(x)=\bigg| \frac{x}{y}\frac{\partial y}{\partial x}\bigg|
\label{eqx3}
\end{equation} 

Under this definition, larger sensitivity means higher severity of fine-tuning. 
Traditionally, a particular value \(c=10\) is chosen as the upper limit for a measurable
parameter to be categorized as ``natural''(or not fine-tuned), although the choice of
 \(c=10\) itself is quite arbitrary. The measure \(c(x)\) along with its cut-off value
constitute the sensitivity criterion.

The sensitivity criterion has been subsequently adopted by many researchers.  
Although the largeness of the sensitivity parameter is usually in good
correspondence with fine-tuning, many researchers soon found it can not
accurately represent the severity of fine-tuning under certain circumstances. 
The most famous examples among them are given by G. Anderson et al\cite{GWA} 
and P. Ciafaloni\cite{CS}.

The example given by  G. Anderson et al\cite{GWA} is regarding the high sensitivity
of proton mass \(m_p\)  to the strong coupling constant \(g\).
Because the relation between the weak scale proton mass \(m_p\), the Planck scale(\(M_P\))
 strong coupling constant \(g\) is:
\begin{equation}
m_p \approx M_Pe^{-(4\pi)^2/bg^2(M_P)}
\label{eqx4}
\end{equation}
which yields the sensitivity:
\begin{equation}
c(g)=\frac{4\pi}{b}\frac{1}{\alpha_s(M_P)}\gtrsim 100
\label{eqx5}
\end{equation}

The large value of \(c\) shows the proton mass is extremely sensitive. According to 
the sensitivity criterion, the proton mass should be highly fine-tuned. But it is well known that
the lightness of the proton mass is the result of the gauge symmetry, not the result
of any delicate fine-tuning mechanisms. Obviously here the sensitivity parameter failed to reflect
the severity of fine-tuning correctly.

The example given by P. Ciafaloni et al\cite{CS} is about the high sensitivity of 
the Z-boson mass. When the Z-boson mass 
\(M_Z\) is dynamically determined through gaugino condensation in a ``hidden'' sector, 
the mass \(M_Z\) can be written as:
\begin{equation}
M_Z\approx M_P e^{-l/g^2_H}
\label{eqx6}
\end{equation}
where \(g_H\) is the hidden sector gauge coupling constant renormalized at \(M_P\).
\(l\) is a constant. Similarly, the sensitivity \(c\) of 
this example is also much greater than the fine-tuning cut-off \(c=10\).
If we follow the sensitivity criterion, 
Z-boson mass \(M_Z\) will be always fine-tuned. This is definitely not true. 
All these examples show that the sensitivity parameter is not a reliable measure of 
fine-tuning.

In order to avoid such misleading results, many authors have attempted to explain these problems, proposed 
alternative prescriptions that supposed to be able to give correct results under these circumstances.
G. Anderson et al~\cite{GWA,AC2,AC3}. first introduced the idea of probability
distribution. They argued that, some physical parameters do have intrinsic large
sensitivity. We should use \(\bar{c}\), the probability average of the 
sensitivity \(c\), to rescale the sensitivity parameter. The result will reflect the
fine-tuning correctly:

\begin{equation}
\gamma = c/\bar{c}
\label{eqx7}
\end{equation}

Under this prescription, only those with \(\gamma \gg 1\) can be categorized as fine-tuned.
This criterion gives correct judgments for the examples discussed above, although they didn't 
explain why some parameters may have intrinsic large sensitivity and why
the rescaling can remove the problem. 

Other authors proposed a modified version of the sensitivity parameter to solve the 
problem~\cite{CS,BR,BS,GRS,RS}:

\begin{equation}
c(x_0)=\bigg| \frac{\Delta x}{y}\frac{\partial y}{\partial x_0}\bigg|
\label{eqx8}
\end{equation} 
where \(\Delta y\) is the experimentally allowed range of the parameter \(x\). They argued
that properly choose the experimentally allowed range \(\Delta x\) can solve these problems. But as 
we know, the fine-tuning problem is an intrinsic property. It should not depend on
any experimental technologies we used to measure a physical quantity. 

Although many authors have attempted to give correct numerical descriptions of fine tuning in
various ways,
yet none of them can claim quantitative rigor. No explanation has ever been proposed
to explain why sometimes we have such large sensitivities for the parameters which are apparently 
not fine-tuned.
It is still unclear how to quantitatively describe the fine-tuning problem in a correctly way. 
The calculated fine-tuning level usually depends on what criterion we use, and how we use it.
Their judgments may reflect the naturalness properties correctly or incorrectly.
Because the sensitivity criterion plays such an important role, it is worth to investigate 
the relationship between the severity of fine-tuning and the sensitivity, find the reason why
the sensitivity is so large for those examples we just discussed.

\section{Why the Sensitivity Criterion Fails}

It is meaningless to directly compare two physical quantities with completely different mass dimensions.  
Convert them to a nondimensionalized format is the most reasonable approach. For physical 
quantities \(x_i(i=1,2,\cdots)\) with different mass dimensions, usually they are first 
nondimensionalized  to \(\delta x_i/x_i\). By doing so it is believed that
the problem of comparing physical quantities with different mass dimensions has been solved.

The examples we discussed in the introduction have many similarities.
For example, Eq.~\ref{eqx4} is almost identical to  Eq.~\ref{eqx6}. 
Both of them are just the simplest renormalization relations
between a dimensionless coupling and a dimensionful mass parameter.
The fact that the sensitivity parameter fails in such simple equations
reminds us that the problem maybe related with the mass dimension. 
The nondimensionalization method we used may be problematic. Some 
effects related with dimensionality haven't been fully removed.
To further investigate the origin of the problem, it is crucial to revisit the relation 
between the mass dimension and the renormalization related fine-tuning problems.

As we know, the scalar mass  \(m\) of Eq.~\ref{eqx2} roughly observes the following
scaling law between the mass and the energy scale parameter \(\Lambda\):

\begin{equation}
m\sim\Lambda^1
\label{eqx9}
\end{equation}
 
If the observable mass \(m\) is 
light, due to the tremendous energy scale
difference between the weak and the Planck scale, precise matching of the initial condition is required.
As a consequence, this type of scaling relation will end up in a fine-tuning problem. 
Unlike the scalar mass,
 the fermion masses are protected by the gauge symmetry, therefore they do not have 
the similar scaling relation. Now suppose we have 
a gauge theory with a dimensionless parameter \(g\) and dimensionful mass parameters \(m_i\).  
Without considering any specific interactions, the lowest order renormalization 
group equations can be written as\cite{Pich,Luo}:

\begin{equation}
\frac{dm_i}{dt}=\gamma_{ij}(g)m_j+\cdots
\label{eqx10}
\end{equation}

\begin{equation}
\frac{dg}{dt}=\beta(g)+\cdots
\label{eqx11}
\end{equation}
where \(t=\ln \Lambda/\Lambda_0\).  \(\beta\) and 
\(\gamma_{ij}\) are dimensionless functions of the coupling constants. For example,
in QCD they are\cite{MV1,MV2,MV3,Luo}:
\begin{equation}
\beta(g)=-b\frac{g^3}{16\pi^2}+...
\label{eqx12}
\end{equation}

\begin{equation}
\gamma_{ij}=\gamma^0_{ij}\frac{g^2}{16\pi^2}+...
\label{eqx13}
\end{equation}
Solving Eq.~\ref{eqx10} and Eq.~\ref{eqx11}, we have:
\begin{equation}
\mathbf{m}=\mathbf{m}_0e^{\int_{t_0}^{t}\mathbf{\Gamma}dt}+\cdots
\label{eqx14}
\end{equation}
where matrices \(\mathbf{m}=(m_i)\), \(\mathbf{\Gamma}=(\gamma_{ij})\).
\(\mathbf{m}_0\) is the initial value of \(\mathbf{m}\) at \(\Lambda_0\).
Its value does not depend on the 
renormalization(renormalization invariant). 
If this model only has one mass parameter \(m\), then Eq.~\ref{eqx14} can be further simplified as\cite{Pich}:
\begin{equation}
m(t)=m(t_0)\Biglb(\frac{g(t)}{g(t_0)}\Biglb)^{-\gamma^0/b}
\label{eqx15}
\end{equation}

If ignore the dependency of  the renormalization invariant quantity \(m(t_0)\) to 
the variation of \(g(t_0)\), the sensitivity 
of the mass parameter to the coupling constant is:

\begin{equation}
c\approx\gamma^0/b
\label{eqx16}
\end{equation}

Because the anomalous dimensions are usually small\cite{pdg}, based on the sensitivity criterion,
the result of Eq.~\ref{eqx16} is certainly not fine-tuned. 

The above descriptions are well-known but they are not complete. In any models,  besides these 
explicit mass parameters,
there always exists an implicit dimensionful parameter : the energy scale parameter \(\Lambda\).
Its effects on  the sensitivity parameter are widely ignored. 
Unlike the other explicit mass parameters,
because in any renormalization group equation,
each term should be dimensionally consistent. This requires that the right hand side of the renormalization
group equation of a dimensionless coupling constant (for example, Eq.~\ref{eqx11})
can not have the first order term.  The consequence of this requirement is that 
the mathematical relation between a dimensionless parameter and  
the energy scale parameter \(\Lambda\) must be an exponential function. This means 
the energy scale parameter 
\(\Lambda\) is always sensitive with respect to minute variations of the related 
dimensionless coupling constants. 
The large sensitivity of \(\Lambda\) here is originated from  different mass dimensions
between a dimensionless coupling and \(\Lambda\), it should not be understood as a fine-tuning problem.  
Because the relation between \(\Lambda\) and \(g\) is nonlinear, 
If for some reason a not-fine-tuned dimensionful parameter is linearly proportional to \(\Lambda\),
then the nondimensionlized method used in the sensitivity parameter will not be able to
remove the effect of different mass dimensions. The large sensitivity of \(\Lambda\) to \(g\) then will
 affect the reliability of the sensitivity criterion.

Generally, from the renormalization point of view, 
a physical mass parameter can be divided into two factors: the factor 
that depends on the running of the energy scale(for example, \((g(t)/g(t_0))^{-\gamma^0/b}\) in Eq.~\ref{eqx15}), and the factor that is renormalization invariant(for 
example, \(m(t_0)\) in Eq.~\ref{eqx15}). Usually because of the protection of the gauge symmetry, 
the one loop correction to the mass parameter will diverge logarithmically rather than quadratically:

\begin{equation}
\delta m \approx g^2\ln{\frac{\Lambda}{\Lambda_0}}
\end{equation}

The logarithm function relieves the high sensitivity of  \(\Lambda\) to \(g\). Therefore the 
renormalization dependent factor of a not-fine-tuned physical mass won't contribute to the sensitivity problem discussed 
in the previous part. 

But it is still possible that a physical mass could be linearly proportional to the energy scale parameter 
in the renormalization invariant factor. If so the sensitivity of the physical mass will be greatly affected
by the high sensitivity of \(\Lambda\) to \(g\). The severity of fine-tuning 
will be overestimated, even though the mass parameter itself is protected by gauge symmetry and is not fine-tuned.  

The renormalization invariant factor of a physical mass  usually is its initial value at 
the a given energy scale.  Therefore principally, any mechanisms that can linearly relate the initial value 
 to the energy scale parameter will cause the problem we discussed in the previous part. For example,
for theories that are asymptotical free, the running of the coupling constant \(g\) can
produce a mass via dynamical symmetry breaking\cite{Gross}:

\begin{equation}
m(g,\Lambda_0)=\Lambda_0e^{-\int dg/\beta{g}}
\label{eqx17}
\end{equation}

The mass defined by Eq.~\ref{eqx17} is certainly highly sensitive to \(g\). 
If the initial value of a physical mass is given by Eq.~\ref{eqx17}, 
then apparently the physical mass will also be highly sensitive to \(g\).
Even if the gauge symmetry stabilizes the mass, prevents quadratic divergent in the
renormalization,  it still can not prevent the physical mass linearly proportional to \(\Lambda\)
in the renormalization invariant factor. The sensitivity of the physical mass to the coupling constant will
be extremely large.

The mechanism of the dynamical symmetry breaking is widely used in many models. For example, 
in supersymmetric standard models, dynamical symmetry breaking is used to specify the soft 
masses\cite{Witten}. As a consequence, any masses related with this mechanism will  be highly 
sensitive to the variations of the dimensionless coupling \(g\). 
the severity of fine-tuning will be greatly over-estimated.
This is the reason why the sensitivity parameter 
fails in Eq.~\ref{eqx4} and Eq.~\ref{eqx6}.

Besides the dynamical symmetry breaking, those mechanisms that end up with a linear relation between a
physical mass and \(\Lambda\) also can cause the same sensitivity problem.
For example, chiral symmetry breaking\cite{Gellmann}, which has the relation:
\begin{equation}
\langle0|q\bar{q}|0\rangle^{1/3}\sim \Lambda
\label{eqx177}
\end{equation}
If the initial value of a nucleon mass is defined by Eq.~\ref{eqx177}, then the nucleon mass will 
 highly sensitive to the variation of \(g\). Other mechanism like gaugino condensation also 
has the same effect.

The problems of the sensitivity parameter have a very close theoretical connection to the problems related with 
the time variations of the fine structure constant.
Recent astrophysical observations have shown many evidences of small time
variation of the gauge coupling constant\cite{Webb}, Many researchers have pointed out that the small time
variations of the fundamental couplings will produce large time variations of various physical parameters, 
such as proton mass and magnetic moment\cite{Calmet,Dent,Langacker}. Therefore, we need to explain 
why these physical quantities can have such large time variations.
Technically, this phenomenon is identical to the problems we just discussed. The time 
variations of the fine structure constant \(\dot{\alpha}\) corresponds to 
the variations of the initial value \(\delta x\) here. Similar to our analysis, 
the large time variations of the physical quantities are caused by the large time variations of 
the energy scale parameter \(\dot{\Lambda}\), and the large time variations of the energy scale
parameter  \(\dot{\Lambda}\) are caused by the small time variations of the fine structure constant 
\(\dot{\alpha}\). The last step is the consequence of the different mass dimensions between 
the fine structure constant \(\alpha\) and the energy scale parameter \(\Lambda\). Based on our analysis,
we can conclude that the large relative changes of nucleon mass and many other dimensionful quantities
in the time-varying coupling constant problem are originated from the  
the native scale difference between the energy scale parameter and the dimensionless coupling constants.

Obviously, these problems only happen when
two physical quantities have different mass dimensions. If two physical quantities 
have the identical mass dimension, then theoretically they won't have the problem.
To verify this, suppose we have two dimensionful mass parameters \(m_i\)  and \(m_j\), 
the solution of Eq.~\ref{eqx10} now becomes\cite{Pich}:

\begin{equation}
m_i(t)=\sum_{j,k}U_{ij}e^{\int dt(-\mathbf{\Gamma}^0/b)}U^{-1}_{jk}m_k(t_0)
\label{eqx20}
\end{equation}
where \(\mathbf{\Gamma}^0\) is the matrix formation 
of \(\gamma^0_{ij}\), \(U_{ij}\) is the element of the matrix that diagonalizes \(\mathbf{\Gamma}^0\).

Because in Eq.~\ref{eqx20},  the exponent \(\mathbf{\Gamma^0}/b\) does not explicitly contain any 
mass parameter, thus the value of \(\partial{m_i(t)}/\partial{m_j(t_0)}\) will not be affected 
by the factor \(\partial{\Lambda}/\partial{g}\). So it is quite straightforward to conclude that 
the sensitivity of a dimensionful mass parameter \(m_i(t)\) to another 
dimensionful mass parameter \(m_j(t_0)\)  won't
 have the same problem we discussed in the previous part.
Similarly, we can also conclude that the problems caused by the time-varying coupling constant 
only exist in the dimensionful parameters. The dimensionless parameters won't have
such problems.

\section{Limitations of the Sensitivity Criterion}

Although the sensitivity criterion has these problems, 
it doesn't mean this criterion will fail under
all circumstances. The problems of the sensitivity criterion occurs only when two parameters
have different mass dimensions, one is a dimensionful quantity, and the other is
a dimensionless coupling constant. They are at the different energy scales and need to be 
mathematically linked by renormalization. Besides, 
there should be a mechanism which makes the renormalization invariant factor of the
dimensionful mass quantity linearly proportional to 
the energy scale parameter \(\Lambda\).

However, the fine-tuning phenomena exist not only in renormalization related problems,
but also in problems such as mass mixing\cite{Casas:2004gh,AD}, where 
all parameters involved in a fine-tuning problem are at the same energy scale and
a renormalization evolution is not required. Therefore
the effects introduced by different mass dimensions have no place to play such a role. The severity
of fine-tuning won't be overestimated if judged by the sensitivity parameter.
So estimate the severity of fine-tuning by comparing two parameters with different 
mass dimensions at different energy scales is the only situation that we need to pay special 
attention to the validity of the sensitivity criterion. 

To solve these problems, many researchers have proposed many alternatives 
methods\cite{GWA,AC2,AC3,CS,BR,BS,GRS,RS}. These methods 
restrict the value of the sensitivity parameter either by
presetting an upper limit to the value of a language parameter or by
rescaling the sensitivity parameter by a background sensitivity. 
Technically these prescriptions do relieve the problem, but they all based
on incorrect theories. Without the 
knowledge of why the sensitivity parameter fails to reflect the severity of fine-tuning,
the results based on these prescriptions could also be misleading.
Certainly, due to the complexity of real models, different
mechanisms could have different types of fine-tuning problems.
It is difficult to invent a simple yet universal prescription 
that can be easily applied to any problems. 
The advantages of the sensitivity criterion is obvious, though it has 
problems under these special situations. If we want to find a way to measure the fine-tuning 
correctly without losing the beauty of simplicity and straightforwardness, it is better 
to keep the sensitivity criterion but avoid using it when two parameters with different 
mass dimensions at different energy scales are compared.
Because this is the simplest and the most efficient solution to measure the naturalness correctly
while avoiding all these problems. Certainly this solution won't restrict our ability to describe the severity 
of the fine-tuning, because the fine-tuning can also be measured by comparing parameters with the identical mass dimension. 

One may argue that, these problems could  be explained in other way.  For instance, 
The sensitivity parameter is correct, while those mechanisms such as dynamical symmetry breaking
are fine-tuned and should be avoided. Because the methods like 
dynamical symmetry breaking are used so widely, if this argument were true, the consequences would be 
disastrous. The astrophysical observations of the time-varying gauge coupling constant give us 
a clue to exam our theory experimentally. Maybe in the future we can find an astrophysical way 
to measure the time variations 
of the proton mass or similar dimensionful physical quantities. By comparing the time variations of a dimensionful
parameter with the time variations of the gauge coupling, we will not only have the experimental evidence to 
verify our conclusion, understand the actual role of the mechanisms like dynamical symmetry breaking, 
but also have the knowledge of the largest sensitivity allowed in our universe, 
not just the arbitrary value \(c_{\text{max}}=10\) chosen by R. Barbieri and G.F. Giudice.

\section{Conclusion}
The problems of the sensitivity parameter has been discovered for more than ten years.
Many analyses, explanations and alternative prescriptions have been proposed to 
solve these problems. But none of them has been accepted widely. The reason why the
sensitivity parameter fails under specific circumstances  still remains unclear.

we investigated the problems exist in the 
sensitivity parameter, demonstrated that the reason why the sensitivity parameter fails to 
represent the true level of fine-tuning is because  the energy scale parameter directly appears in 
the renormalization invariant factor of a dimensionful parameter via mechanisms other than 
the renormalization. Thus the large sensitivity of  the energy scale parameter to the 
dimensionless coupling will greatly influence the sensitivity of the dimensionful parameter, 
severity of fine-tuning will be over-estimated. Theoretically, the problems of the sensitivity 
parameter are similar to the problems related with time-varying coupling constant. Results in 
that field can be applied in the sensitivity research.

Finally, we want to point out that this effect only exists when parameters with different 
mass dimensions at different energy scales are compared. Based on these analyses, we argue that
the best way to avoid these problems is always compare parameters with same mass dimensions
if they are at the different energy scales.

\end{document}